\documentclass[conference]{IEEEtran}
\IEEEoverridecommandlockouts
\usepackage{cite}
\usepackage{amsmath,amssymb,amsfonts}
\usepackage{algorithmic}
\usepackage{graphicx}
\usepackage{textcomp}
\usepackage{xcolor}
\usepackage{url}
\usepackage{balance}
\def\BibTeX{{\rm B\kern-.05em{\sc i\kern-.025em b}\kern-.08em
    T\kern-.1667em\lower.7ex\hbox{E}\kern-.125emX}}
\begin{document}

\title{ABET Accreditation: A Way Forward for PDC Education}

\author{ 
\IEEEauthorblockN{\makebox[0.5\columnwidth]{Sherif G. Aly}}
\IEEEauthorblockA{\textit{Department of Computer Science and Engineering} \\
\textit{The American University in Cairo}\\
Cairo, Egypt \\
sgamal@aucegypt.edu}
\and 
\IEEEauthorblockN{ \makebox[0.5\linewidth]{Haidar Harmanani}}
\IEEEauthorblockA{\textit{Department of Computer Science and Mathematics} \\
\textit{Lebanese American University}\\
Beirut, Lebanon \\
haidar@acm.org}
\and
\IEEEauthorblockN{ \makebox[0.5\linewidth]{Rajendra K. Raj}}
\IEEEauthorblockA{\textit{Department of Computer Science} \\
\textit{Rochester Institute of Technology}\\
Rochester, NY, USA \\
rkr@cs.rit.edu}
\and 
\IEEEauthorblockN{Sanaa Sharafeddine}
\IEEEauthorblockA{\textit{Department of Computer Science and Mathematics} \\
\textit{Lebanese American University}\\
Beirut, Lebanon\\
sanaa.sharafeddine@lau.edu.lb}
}

\maketitle

\begin{abstract}

With parallel and distributed computing (PDC) now wide-spread, modern computing programs must incorporate PDC within the curriculum. ACM and IEEE Computer Society's Computer Science curricular guidelines have recommended exposure to PDC concepts since 2013. More recently, a variety of initiatives have made PDC curricular content, lectures, and labs freely available for undergraduate computer science programs. Despite these efforts, progress in ensuring computer science students graduate with sufficient PDC exposure has been uneven.

This paper discusses the impact of ABET's revised criteria that have required exposure to PDC to achieve accreditation for computer science programs since 2018. The authors reviewed 20 top ABET-accredited computer science programs and analyzed how they covered the required PDC components in their curricula. Using their own institutions as case studies, the authors examine in detail how three different ABET-accredited computer science programs covered PDC using different approaches, yet meeting the PDC requirements of these ABET criteria. The paper also shows how ACM/IEEE Computer Society curricular guidelines for computer engineering and software engineering programs, along with ABET accreditation criteria, can cover PDC.

\end{abstract}


\section{Introduction}
\label{INTRO}

Parallel computing is often traced to the {\em D825} multiprocessor system in the 1960s ~\cite{D825_1962}, and distributed systems to early computer networks and client-server computing in the 1970s \cite{tanenbaum2016}. Concurrent programming also became commonplace around the same time-frame, with synchronizing communicating processes formalized by Disjktra~\cite{THE} and Hoare~\cite{CSP}. Nowadays, much of the dramatic advances in a wide range of disciplines are enabled by parallel and distributed computing (PDC). For example, Internet of Things (IoT) provides the basis for computing devices of heterogeneous nature, technology, and capability to cooperate and deliver added value services across diverse industries. Mobile healthcare, industrial automation, holographic teleportation, autonomous vehicles, smart environments, digital currency mining, and GPU computing in general are examples of systems enabled by revolutionary advances in communications and parallel computing~\cite{9145564}.  Such systems are founded by services that inherently need parallel processing, inter-service communication, and distributed monitoring and control. 

 Given the state of today's computing, it is crucial that computing professionals understand PDC intimately, and for computing education programs to incorporate PDC concepts, not just at the graduate level but also at the undergraduate level. Although awareness of the need for PDC education has existed for more than a decade,  computing education has been slow at formalizing this need and making it actionable. In the case of the dominant Computer Science (CS) discipline, the ACM and IEEE Computer Society took some initial steps in the CS2013~\cite{CS2013} curricular guidelines to highlight the importance of PDC as one of the core tier-1 and tier-2 areas. Moreover, the guidelines expected additional coverage of PDC concepts intrinsic to operating systems, computer architecture, database management systems, and computer networks. More recently, the visionary CC2020 described draft PDC competencies for CS~\cite[pp. 112--123]{CC2020final}.

Despite these developments, many computer science programs have not paid sufficient attention to including PDC in their curricula, and ultimately preparing graduates to meet today's needs despite the availability of textbooks~\cite{kirk,quinn}, as well as extensive curricular resources developed by the Center for Parallel and Distributed Curriculum Development and Educational Resources (CDER)~\cite{CDER},  PDC~Unplugged~\cite{PDCUnplugged}, and researchers \cite{PrasadText,Fernandez2019,Olivera2019,Homma2008:HPC,ToUCH}.

Recognizing these professional needs, as well as the lack of sufficient incentives, ABET's~\cite{ABET} Computing Accreditation Commission (CAC) revised its criteria for undergraduate Computer Science programs by requiring coverage of PDC concepts~\cite{CACCriteria}. These criteria were first released in 2018, and since 2019, all ABET-accredited CS programs are required to include PDC in their curricula. As ABET typically follows a six-year review cycle, it is likely that all [currently] 398 ABET-accredited CS programs will have coverage of PDC content by 2024. 

This paper explores the impact of the revised ABET's CS Program Criteria on the incorporation of  PDC content into modern ABET-accredited computer science programs. Section~\ref{Role} discusses the history and formulation of the PDC requirements in the ABET CAC Criteria, along with approaches to covering PDC not only to meet the ABET CAC Criteria requirements, but also to enhance them. Section~\ref{survey} presents the results of a survey of top ABET-accredited CS programs to understand how they support PDC concepts across their curricula. Using three examples of ABET-accredited programs, Section~\ref{CS CASE STUDIES} expands on actual implementation of the different approaches to satisfy the PDC requirements of the CAC Criteria. Section~\ref{PDCENG} explores whether the ABET engineering criteria~\cite{EACCriteria}, along with the ACM/IEEE Computer Society curricular guidelines for computer engineering~\cite{ACM_CE} and software engineering~\cite{ACM_SE}, essentially require coverage of PDC concepts to ensure that graduates are prepared to enter engineering practice. The paper concludes with remarks about how PDC coverage can be improved in computing and engineering programs.
\section{Using Accreditation to Move PDC Forward}
\label{Role}

The world of accreditation differentiates between {\em institutional accreditation} where regional or national accreditation organizations provide assurance that a university as a whole reliably provides quality education and {\em program accreditation} where accrediting bodies use disciplinary guidelines to establish and verify the quality of the degree program and increasingly, the characteristics of the program graduates~\cite{Oudshoorn2018:ASEE}. Program-level accreditation focuses on areas such as student advising, graduation requirements, faculty qualifications, curriculum standards, and available facilities. In the US (and over 40 other countries), ABET~\cite{ABET} has been the accrediting body for programs in computing and engineering, through its Computing Accreditation Commission (CAC) and Engineering Accreditation Commission (EAC). The rest of this section examines PDC requirements in the CAC Criteria, specifically Computer Science; Section~\ref{PDCENG} explores PDC in Software Engineering and Computer Engineering.

\subsection{ABET's Computer Science Accreditation Criteria}
\label{CS Criteria}

ABET's program criteria across all its commissions are created by the professional society representing the corresponding discipline. For instance, for computing disciplines including computer science, the professional society is CSAB, the joint body created by ACM and IEEE Computer Society~\cite{CSAB}. As mentioned earlier, the ACM and IEEE Computer Society have jointly developed different curricular guidelines for a variety of computing disciplines. In the case of CS, the 2013 ACM/IEEE Computer Society's Computer Science Curricular Guidelines (CS2013)~\cite{CS2013} have been the latest. These guidelines highlighted the importance of parallel and distributed computing and noted that ``parallel and distributed computing has moved from a largely elective topic to become more of a core component of undergraduate computing curricula."

Therefore, the accreditation criteria for CS developed by ABET draw substantially from CS2013~\cite{CS2013}, which defines parallel and distributed computing to encompass the following~\cite{CS2013}:

\begin{enumerate}
    \item  An understanding of fundamental systems concepts such as concurrency and parallel execution, consistency in state/memory manipulation, and latency.
    
     \item Understanding of parallel algorithms, strategies for problem decomposition, system architecture, detailed implementation strategies, and performance analysis and tuning.
    
    \item Message-passing and shared-memory models of computing.
\end{enumerate}

CC2020~\cite{CC2020final} reiterates the above knowledge areas and recommends specific topics including a coverage of a parallel divide-and-conquer algorithm, critical path, race conditions, processes, deadlocks, and properly synchronized queues.

What follows  examines how PDC is required by the CAC Criteria for accrediting Computer Science programs and what approaches can be used by accredited programs to meet this requirement. 

Program-level accreditation of Computer Science (CS) programs has been ongoing since the 1985-86 academic year.  At present, the CAC accredits 398 CS programs around the globe using the {\em ABET Criteria for Accrediting Computing Programs (Criteria)}~\cite{CACCriteria}. The CS Criteria comprise the {\em General Criteria} that apply to all computing programs and the {\em Computer Science Program Criteria} that are specific to CS programs.  

\begin{figure}[tb]
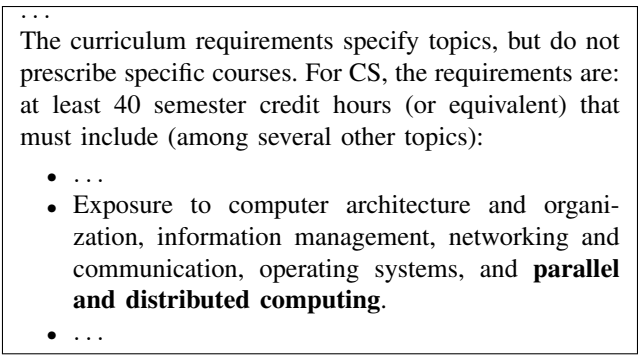

\centering
\fbox{
\parbox{0.9\columnwidth}{
$\ldots$\\
The curriculum requirements specify topics, but do not prescribe specific courses.  For CS, the requirements are: at least 40 semester credit hours (or equivalent) that must include (among several other topics):
\begin{itemize}
\item $\ldots$
\item Exposure to computer architecture and organization, information management, networking and communication, operating systems, and {\bf parallel and distributed computing}.
\item $\ldots$
\end{itemize}
}
}
\caption{Computer Science Program Criteria---Curriculum}
\label{fig:cs criteria}
\end{figure}

This paper focuses on the CS programs because they are now required to cover PDC by the corresponding criteria, within the curriculum requirements, as depicted in Fig.~\ref{fig:cs criteria}. In other words, in line with the CS2013 curricular requirements~\cite{CS2013}, the ABET CS Program Criteria require exposure to computer architecture and organization, information management, networking and communications, operating systems, and parallel and distributed computing. The criteria explicitly point out that these requirements are flexible, and do not necessarily ask for courses to be introduced into programs, but rather topics or knowledge areas that ought to be covered somewhere in the program requirements. For example, some topics may be covered in a relatively small number of lectures embedded within an existing course. 

The phrase ``exposure to" used in these criteria is in line with the CS2013 curricular requirements. With the anticipated  continued growth of PDC in computing, the authors assume the next decennial revision of the ACM CS guidelines will emphasize PDC even more. In the meantime, as discussed in the next two sections, many excellent programs in CS are already emphasizing PDC, even beyond  what the ABET CS Criteria require.


 \subsection{Meeting the PDC Exposure Requirement in the Computer Science Criteria}
\label{Meeting CS Criteria}

The Computer Science Criteria are flexible and provide several avenues for the PDC content to be included. Opportunities thus
exist across various courses within a Computer Science curriculum to include PDC exposure. For the accreditation criteria to be satisfied, a program must include PDC coverage in required coursework to ensure all  graduating students receive the required exposure. Based on curricula formulated by CDER~\cite{CDER}, three core PDC concepts have been clearly identified namely, {\em concurrency}, {\em parallelism}, and {\em distribution}~\cite{Raj2020HighPC}.
These topics can be covered in courses such as computer programming, algorithms design, operating systems, databases, networking, and architecture.

Newhall~\cite{NEWHALL201753} explored how PDC could be included when planning or revising the curriculum:
\begin{enumerate}

\item Achieve an early exposure and maturity of PDC topics to allow learners an opportunity to achieve further in-depth knowledge where needed before graduation.
\item Design a curriculum that intentionally overlaps some topics across the curriculum.
\item Provide breadth of knowledge as well as appropriate in-depth coverage.
\item Expose learners to the topics in multiple sub-disciplines as opposed to coursework that only covers PDC topics.
\end{enumerate}

Most programs follow one or both of two major approaches to incorporating any content in curricula. In the first approach, programs dedicate at least one required course to cover the required knowledge units for any required area. For PDC, such a required course would need to contain essential parallel programming concepts, as well as more advanced parallelism and distribution concepts. This content can be developed following the guidance provided by CS2013~\cite{CS2013} or the draft PDC competencies mentioned in CC2020~\cite{CC2020final}. 

The dedicated-course approach is the obvious approach, as it provides faculty and students with the opportunity to explore a wide array of PDC topics while emphasizing others that may have been offered in other courses in the program. A course like this could tap into the vast amount of resources that are made available by Intel, NVIDIA, and IBM, among others, which include course materials, textbooks, free compilers, free training, and in some instances free hardware. Also, CDER~\cite{CDER} and PDC Unplugged~\cite{PDCUnplugged} provide a multitude of curricular content and labs that can be easily included.
    
On the other hand, due to limits imposed by faculty availability and expertise, available number of course credits, and related constraints, a program may find that it is simply not possible to add a dedicated course to the program curriculum. However, as PDC concepts are inherent in various computing areas, it is not hard to integrate different parts of the knowledge area into existing courses in the curriculum, rather than create a dedicated course. Thus, PDC coverage is supported at the periphery of the curriculum by scattering the knowledge units into various required courses. Some programs would probably find this approach more practical to satisfy accreditation requirements while providing the students with some exposure to PDC.  Table~\ref{topics} shows how a typical accredited computer science program may choose to include coverage of PDC across different courses such as Systems Programming, Computer Organization or Architecture, Operating Systems, and Database Systems.

\begin{table*}[tb]
\caption{Mapping Different PDC Concepts to Typical Courses }
\label{topics}
\begin{center}
\begin{tabular}{|l|p{1.5cm}|p{1.6cm}|p{1.2cm}|p{1.1cm}|p{1.1cm}|}
\hline
& Systems Programming & Computer Organization/ Architecture & Operating Systems & Database Systems & Computer Networks\\\hline
Programming with threads & $\times$ & & $\times$ & & $\times$\\ \hline
Transactions processing & & & &$\times$ &\\ \hline
Parallelism and concurrency & $\times$ & $\times$ & $\times$ & $\times$&$\times$\\ \hline
Shared-Memory programming & $\times$ & & $\times$ & & \\ \hline
Inter-Process Communication (IPC)& $\times$ & & $\times$ & & $\times$\\ \hline
Atomicity & $\times$ & & $\times$ & &\\ \hline
Performance measurement, speed-up, and scalability &  & $\times$ & & &\\ \hline
Multicore processors & & $\times$ &  & &\\ \hline
Shared vs. distributed memory & $\times$ & $\times$ & $\times$ &  &\\ \hline
SIMD and vector processors & &  $\times$&  & &\\ \hline
Instruction Level Parallelism & & $\times$& & &\\ \hline
Flynn’s taxonomy & & $\times$ & & & \\ \hline
Client-server programming & $\times$ &  & & &$\times$\\ \hline
Memory and caching  & $\times$ & $\times$ & $\times$ & &\\ \hline
\end{tabular}
\end{center}
\end{table*}%
\section{PDC in 20 Top Accredited CS Programs}
\label{survey}

To  understand how ABET-accredited programs have incorporated PDC in their curriculum, the authors collected data from 20 ABET-accredited computer science programs that are ranked in the top $100$ by US News~\cite{USNews}.

\begin{figure*}[tb]
\centerline{\includegraphics[width=1.25\columnwidth]{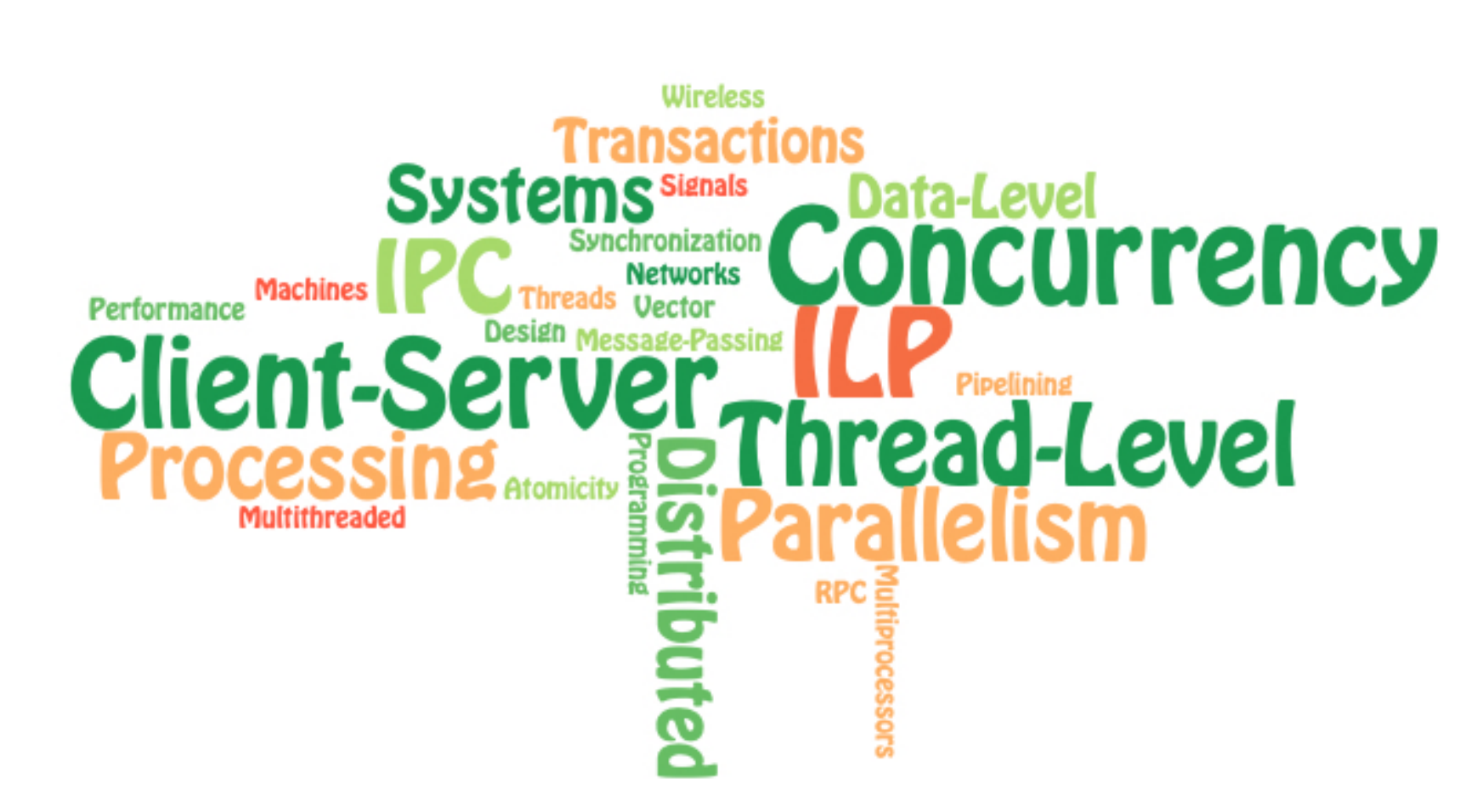}}
\caption{PDC Topics Used by Surveyed Programs for ABET Accreditation}
\label{fig:topics}
\end{figure*}

\begin{figure*}[bt]
\centerline{\includegraphics[width=1.25\columnwidth]{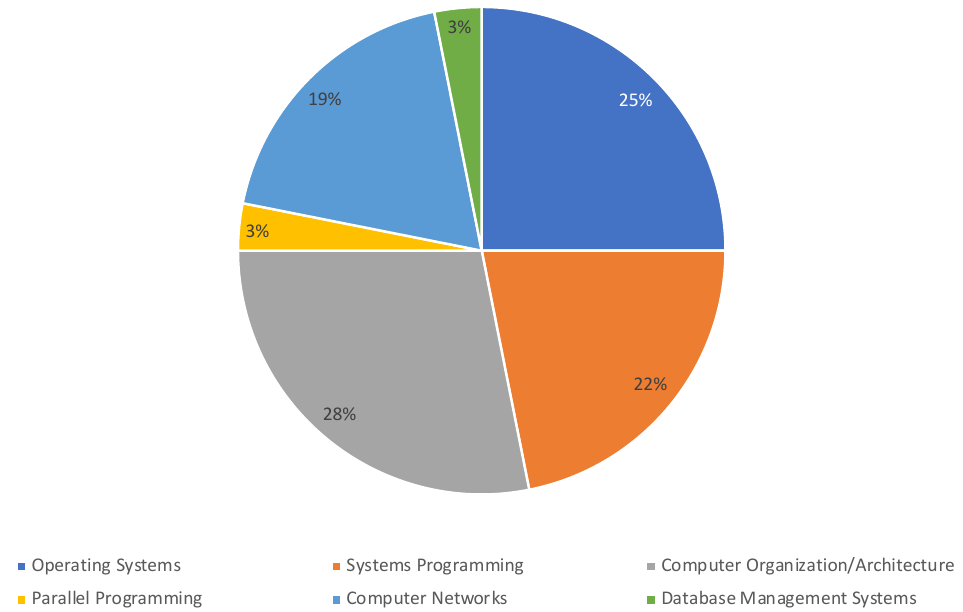}}
\caption{Courses for PDC Content by Surveyed Programs for ABET Accreditation}
\label{fig:courses}
\end{figure*}

The collected data was studied with a focus on required courses that included PDC components.  The course descriptions and the learning outcomes were analyzed using information that were published online.  A weighted sum of all courses that tackle specific components of the PDC knowledge area was computed. Fig.~\ref{fig:topics} shows  common topics covered by the surveyed programs that help to support PDC for ABET Accreditation.  Fig.~\ref{fig:courses} shows the percentage of courses that encompass these topics.  It is worth noting that out of the 20 surveyed programs, only one program had a dedicated parallel programming course while the remaining programs used multiple courses to cover PDC topics.

Most modern CS programs offer the following courses, several of which are required:
\begin{enumerate}
    \item A typical {\em operating systems} or {\em systems programming} course can include coverage of concurrency, atomicity, shared memory programming, and inter-process communication.  Programming with threads can also be covered in a more elaborate way using client-server programming in a computer networks course or in systems programming course;
    \item A {\em database management} course can incorporate distributed computing concepts including transactions processing, scheduling concurrent transactions, transactions locks, and deadlocks;
    \item A {\em computer organization} or {\em architecture} course can incorporate Amdahl’s law and its implication on the performance of a particular parallel algorithm, speed-up and scalability, Flynn's taxonomy, the concept of multicore processors, multiprocessor caches and cache coherence, pipelining, instruction level parallelism, and vector and SIMD computers. Advanced computer architecture courses can also cover concepts of shared and distributed memory;
    \item {\em Programming language} courses can incorporate support for parallel and distributed computing concepts, including but not limited to, programming with threads, programming language constructs and APIs for thread creation and management, networking to support distributed computing, inter-process communication and synchronization, along with virtual machine and translator support for parallelization;
    \item Selected parallel algorithms and related theoretical analysis and applications can be covered in a {\em design and analysis of algorithms} course;
    \item A course in {\em software engineering} can provide a coverage of modeling and notations used for parallel behavior as well as design considerations of distributed components. 
\end{enumerate}

These results show--at a high level--that the ABET Computer Science criteria have already begun to have their anticipated impact on incorporating PDC concepts in undergraduate CS programs. Three different ABET-accredited CS programs on three different continents are now examined for achieving PDC coverage.
\section{Case Studies: PDC in CS Programs}
\label{CS CASE STUDIES}

This section examines three ABET-accredited computer science programs from the authors' own institutions to study the actual coverage of PDC topics.

\subsection{Case Study: Lebanese American University}

The Computer Science Program at the Lebanese American University incorporated parallel programming early in 1996 in the form of a fully-fledged course based on the Network of Workstations (NOW) model~\cite{now,Adams2015:budget:beowulfs} using the Message-Passing Interface (MPI).  The course content was later updated to include GPGPU manycore programming and a brief introduction to deep learning as a case-study to showcase the power of parallelism~\cite{CS-LAU}. It should be noted that in addition to this required course, students explore PDC concepts in various required courses including {\em operating systems}, {\em computer organization}, and {\em database management systems}.

\subsubsection{Meeting the PDC Program Criteria}

The dedicated course introduces relevant parallel architectural trends and aspects of multicores in addition to the design and analysis of parallel programs.  The course description is:
\begin{quote}
This course provides an introduction to parallel programming with a focus on multicore architectures and cluster programming techniques. Topics include relevant architectural trends and aspects of multicores, writing multicore programs and extracting data parallelism using vectors and SIMD, thread-level parallelism, task-based parallelism, efficient synchronization, program profiling, and performance tuning.  Message-passing cluster-based parallel computing is also introduced.  The course includes several programming assignments to provide students first-hand experience with programming, and experimentally analyzing and tuning parallel software.   
\end{quote}

The course learning outcomes are that students will:

\begin{enumerate}
    \item Understand the challenges of as well as the motivations for using parallel programming.
    
    \item Demonstrate an ability to analyze the efficiency of a given parallel algorithm.
    
    \item Demonstrate an ability to design, analyze, and implement programming applications using multicore and manycore systems.
\end{enumerate}

In addition, students need to complete labs, a milestone project in each part, one written report per milestone, and to make one oral presentation. As the course is being used to assess communication skills, all material is graded for grammar, spelling, and style.

The course delivery is easily broken down into three parts.  The first part introduces the history of PDC and the driving forces behind its recent resurrection, tackling issues such as the continuous decrease of feature size, the increase in the number of transistors per chip, clocking, performance scalability, excessive power consumption, and the rise of multicore processors.

The second part of the course focuses mostly on multi-core programming using {\tt Pthreads} and {\tt OpenMP}.  Students are introduced to shared-memory issues including thread-level parallelism, task-based parallelism, data races, false sharing, memory contention, synchronization, and extracting data parallelism using vectors and SIMD.  This part of the course is supported by {\em Intel}, which provides hands-on labs as well as free licenses for the {\em Intel Parallel Studio} compiler.

In the third part, which is roughly 60\% of the course, students are introduced to the concept of manycores focusing on single instruction,  multiple threads (SIMT) execution model that is inherent in GPU devices.  In this part, students develop manycore applications using {\tt CUDA C} and {\tt OpenACC}, and learn advanced memory management techniques as well as using concurrent streams.  This part of the course is supported by NVIDIA's Deep Learning Institute (DLI) which provides all students access to the NVIDIA's online training platforms.  Typically, students earn two NVIDIA certificates in this part, an OpenACC and a CUDA C.  Labs are completed on the cloud using NVIDIA's accelerated computing environment.

The parallel programming course  is also a cornerstone for meeting the ABET criteria.  In fact, the course is used to meet multiple performance criteria in ABET's Student Outcome 2 (Design, implement, and evaluate a computing-based solution to meet a given set of computing requirements in the context of the program’s discipline) and Student Outcome 3 (Communicate effectively in a variety of professional contexts), as well as the curricular requirement for PDC exposure.

\subsection{Case Study: The American University in Cairo}
\label{CS AUC}

The Computer Science and Engineering Department at the American University in Cairo offers two ABET accredited programs: one in Computer Science, and another in Computer Engineering. Both programs follow an early maturity pedagogical approach to ensure students learn key knowledge units as early as possible in the curriculum. The intent is to provide students with an ability to innovate faster, and create more meaningful contributions before graduation. 

The CS program does not require a dedicated course that covers PDC topics, yet the knowledge units to support this requirement are satisfied across various other courses. 
\begin{enumerate}

\item In the fundamental course sequences, students are exposed to a basic ability to create and control threads, as well as a simple client server connectivity. 

\item In the courses of computer organization, and architecture, multiprocessing, thread level parallelism, pipelining, instruction level parallelism, superscalar architectures, VLIW architectures and architectures based on dynamic scheduling such a the non-speculative and the speculative versions of Tomasulo's architectures are covered. 

\item The operating systems course provides coverage at a substantial depth for multi-threading, speedup, multiprocessing, mutual exclusion, synchronization, deadline and starvation, and scheduling on single and multiprocessor systems. 

\item In software engineering, students are exposed to the development of semi-complex systems that involve distribution of components and some parallel functionality and  distribution of components.

\item In concepts of programming languages, design considerations for various programming languages are discussed, including support for networking, and threading. Students work on a project using one of the trending languages, and must present many of the language features including language support for  parallelism and networking, and the effect of garbage collection on currently executing processes and threads.

\item Even though there is a course dedicated to distributed systems within the department, it is only required at the moment for the Computer Engineering program. This fundamentals of distributed computing course covers topics ranging from modeling and specification to consistency and inter-process communication, load balancing, process migration, and distributed challenges. A demanding project helps students master some of the fundamental and advanced topics of the domain.
\end{enumerate}

\subsection{Case Study: Rochester Institute of Technology}
\label{CS RIT}

In 2010, i.e., almost a decade prior to the CAC Criteria requirement for PDC to be covered, the faculty in RIT's computer science program recognized that few modern computers consisted of a single box with a single CPU, and that modern computers typically include multi-core CPUs forming clusters of machines connected by a local network, or are units of distributed systems interconnected over the Internet. The faculty decided then that all computer science students needed to know not only how these these parallel and distributed systems function but also how to write programs for these computing systems. To address this goal, it was decided to create a coherent single semester-long course that would allow students learn the foundations of PDC including principles of parallel computing, basic network protocols, principles of network security, basic architectures of cluster and grid computing, and build additional depth in multithreaded programming and techniques for developing network applications. 

Another important consideration was that these topics needed to be covered in a single course, so that students could understand the interrelationships between these topics, such as the synergies between multithreaded programming and network programming. The focus was not to cover these topics in depth, but instead to provide the breadth of knowledge and skills in PDC for each computer science student. The intention was that students would be ready to take advanced electives in these topics, which could include operating systems, networking, systems programming, distributed systems, parallel computing, grid computing, as well as cryptography and cybersecurity. Before this change, the program had required courses in operating systems and in networking. Obviously after the change, modified courses in operating systems and networking were created as electives to explore advanced topics in greater depth.

The course, \textit{Concepts of Parallel and Distributed Systems}, has been taught since Fall 2013 when RIT transitioned to a semester system. The course description is:
\begin{quote}
    This course is an introduction to the organization and programming of systems comprising multiple computers. Topics include the organization of multicore computers, parallel computer clusters, computing grids, client-server systems, and peer-to-peer systems; computer networks and network protocols; network security; multithreaded programming; and network programming. Programming projects will be required. 
\end{quote}
The course learning outcomes are that students will:
\begin{enumerate}
    \item Explain the concepts of processes, threads, and scheduling.
    \item Develop multithreaded programs.
    \item Explain the concepts of computer networking, the layered network architecture, network security, and network communication with connections and datagrams.
    \item Develop network application programs.
    \item Explain the concepts of distributed system architectures and middleware.
    \item Explain the concepts of parallel computer architectures.
\end{enumerate}
The main topics covered in the course include the following: multithreaded computing (including processes, threads, scheduling, synchronization, deadlock, starvation. and multicore computers); networked computers (including client-server, connections, application protocol design, and socket and datagram programming); network protocols and security;  distributed systems (including different architectures, middleware, distributed objects, and web services), and parallel computing (including architectures and middleware).

In addition to this required course, students explore PDC concepts in earlier courses, starting in their second required programming course in the freshman year, where Java threads and synchronization are explored in depth. The required course, Mechanics of Programming, covers pthreads in depth. Another required course, Concepts of Computer Systems, also covers pipelining and other related PDC concepts.

\section{PDC in Computer and Software Engineering Programs}
\label{PDCENG}

ABET's engineering accreditation criteria~\cite{EACCriteria} for computer engineering (CE) and software engineering (SE)  programs do not explicitly mention PDC. They, however, do require the program curriculum to ``provide adequate content for each area, consistent with the student outcomes and program educational objectives, to ensure that students are prepared to enter the practice of engineering"~\cite{EACCriteria}. Modern curricula for these engineering programs are typically guided by the curricular guidelines developed by ACM and IEEE Computer Society's Computer Engineering~\cite{ACM_CE} and Software Engineering~\cite{ACM_SE}. In essence, any modern computer engineering or software engineering program that lives up to these curricular guidelines will need to cover PDC concepts. In other words, it is likely that ABET's engineering criteria, being based on the ACM/IEEE Computer Society curricular guidelines, require the PDC coverage through courses that are inherently required for both programs.


The computer engineering curriculum guidelines (CE2016) delineate twelve broad knowledge areas for the relevant body of knowledge~\cite{ACM_CE}. Knowledge areas are decomposed into associated knowledge units with a list of learning outcomes each. Knowledge units are classified as core and supplementary units. Core units of each knowledge area constitute its minimal required knowledge and skills that have to be covered within a program curriculum. Computing algorithms, computer architecture and organization, systems resource management, and software design represent knowledge areas of computer engineering that explicitly address parallel and distributed computing concepts through core knowledge units. 

Table~\ref{tab:CE_PDC} lists the knowledge areas that cover PDC-related core knowledge units. As per CE2016, most of the pertinent skills associated with the latter knowledge units are expected to be minimally met with relatively basic ability, while a few of them are set to a higher level of accomplishment.

\begin{table}[tbh]\centering
\caption{PDC in Computer Engineering Knowledge Areas~\cite{ACM_CE}}
\label{tab:CE_PDC}
\begin{tabular}{|p{0.30\columnwidth}|p{0.55\columnwidth}|}
\hline
\textbf{Knowledge Area} & \textbf{PDC-related Core Knowledge Units}\\ \hline
      
Computing Algorithms & Parallel algorithms/threading \\  \hline
        
Architecture and Organization & Multi/Many-core architectures \\
      & Distributed system architectures \\ \hline
      
Systems Resource Management & Concurrent processing support\\ \hline
      
Software Design & Event-driven and concurrent programming \\ \hline
    
\end{tabular}
\end{table}

Similar to CE2016, the software engineering curriculum report SE2014 defines a body of core knowledge that represents the knowledge and skills for which there is consensus that any software engineering graduate should acquire~\cite{ACM_SE}. This core knowledge is coined as software engineering education knowledge areas and referred to as SEEK. SEEK  comprises 10 knowledge areas with a set of associated knowledge units, some of which are designated as essential in relevance to the core knowledge. CE2016 defines a collection of topics that include the cognitive skill level at which each topic of a given knowledge unit is expected to be attained. Three cognitive skill levels are defined with application being the  highest level. 

As shown in Table~\ref{tab:SE_PDC}, computing essentials area is a main knowledge area of SEEK that entails the foundation for the design and construction of software and includes one unit on construction technologies. The former unit emphasizes two PDC-related topics, namely concurrency primitives and construction methods for distributed software. Both topics are classified as essential to the core and expected to be met at the application level.  

\begin{table}[tbh]\centering
\caption{PDC in Software Engineering Knowledge Areas~\cite{ACM_SE}}
\label{tab:SE_PDC}
\begin{tabular}{|p{0.23\columnwidth}|p{0.65\columnwidth}|}
\hline
\textbf{Knowledge Area} & \textbf{PDC-related Core Topics}\\ \hline

Computing Essentials
       & Concurrency primitives (e.g., semaphores and monitors) \\
      & Construction methods for distributed software (e.g., cloud and mobile computing) \\ \hline
\end{tabular}
\end{table}

With reference to the above, any computer engineering or software engineering program addresses concepts on parallel and distributed software with varying levels of attainment. Students understand the intrinsic performance gains of such software, albeit they are aware that the potential gain is highly dependent on how its different parts are set to interact among each other in terms of dependencies, communications, and competitions. Per ACM/IEEE Computer Society curricular guidelines, SE and CE programs do equip their students with tools to combat the challenges of concurrency. Parallel and distributed software can indeed apply to the complex software-related ABET program criteria for CE and SE. ABET program criteria for computer engineering require that students learn to analyze and design complex software among others. Although analysis and design of application-specific algorithms represent one individual knowledge unit of computing algorithms area in computer engineering, incorporating PDC concepts described in both knowledge areas on computing and software design yields complex software, thereby meeting the corresponding ABET program criterion. 

With reference to the ABET program criteria for software engineering, the curriculum must include software engineering tools appropriate for the development of complex software systems. The attainment of the specified PDC-related topics on concurrency and construction methods for distributed software as per SE2014 with the recommended cognitive skill level allows strengthening the compliance with this criterion.

In short, even though the ABET program criteria for computer and software engineering do not explicitly require PDC coverage, it is clear by looking at the curricular guidelines for both programs, it is simply not possible for program graduates in these disciplines to graduate without learning PDC. The computer engineering and software engineering programs at the authors' institutions anecdotally verify this claim.

\section{Final Remarks}
\label{FINAL}

Given the ubiquity of PDC in all aspects of modern society, it is incumbent upon contemporary computing and engineering programs to include appropriate coverage of these topics. Accordingly, parallel and distributed computing knowledge areas are emphasized in the latest ACM curricula, notably CC2020~\cite{CC2020final}'s draft competencies for PDC within CS.  

Program accreditation also has a role to play in moving curricula forward.  The latest ABET's CAC criteria started requiring PDC exposure for CS programs in 2018~\cite{CACCriteria}.  These criteria already have led to systematically increasing PDC coverage, as illustrated by our survey of 20 of the top accredited CS programs and the detailed examination of how three accredited CS programs cover PDC. Two approaches to cover PDC were observed: a predominant one that uses multiple topics across various computer science courses, and a second approach that uses a dedicated PDC course.  Both approaches are viable and meet the current ABET criteria.  

It is expected that PDC will continue to gain importance in computing curricula  with recent computing development and applications.  Engineering, on the other hand, has a more general approach in line with the centuries of engineering education and practice.  Thus, this paper also explored whether today's engineering graduates are also exposed to PDC concepts to be ready for careers in their disciplines. The conclusion is that ABET program accreditation has played a useful role in infusing PDC topics into CS programs around the world at a faster pace, thus leading to stronger and wider PDC coverage.


\section*{Acknowledgment}

Raj acknowledges support provided by the US National Science Foundation under Awards 1922169 and 2021287.

\balance
\bibliographystyle{ieeetr}
\bibliography{IEEEabrv,edupar2021}
\end{document}